\documentclass[superscriptaddress,floatfix,prl,twocolumn,amsmath,amssymb]{revtex4-2}

\usepackage{graphicx}
\usepackage{xcolor}
\usepackage{subfigure}
\usepackage{figsize}
\usepackage{amsmath}
\usepackage{hyperref}
\usepackage{lipsum}
\usepackage{tabularx}
\usepackage{multirow}

\graphicspath{{Figures/}}
\newcommand{\sivminus}{$^{29}$SiV$^{-}~$}
\newcommand{\siv}{$^{29}$SiV$^{0}~$}
\newcommand{\gevminus}{$^{73}$GeV$^{-}~$}
\newcommand{\gev}{$^{73}$GeV$^{0}~$}
\newcommand{\snseven}{$^{117}$SnV$^{0}~$}
\newcommand{\snsevenminus}{$^{117}$SnV$^{-}~$}
\newcommand{\snnine}{$^{119}$SnV$^{0}~$}
\newcommand{\snnineminus}{$^{119}$SnV$^{-}~$}
\newcommand{\cthirteen}{$^{13}$C}

\begin{document}
\title{All-electron $\mathrm{\textit{ab-initio}}$ hyperfine coupling of Si-, Ge- and Sn-vacancy defects in diamond}

\author {Akib Karim}
\email[Author to whom correspondence should be addressed: ]{akib.karim@rmit.edu.au}
\affiliation{Quantum Photonics Laboratory and Centre for Quantum Computation and Communication Technology, School of Engineering, RMIT University, Melbourne, VIC 3000, Australia}
\author {Harish H. Vallabhapurapu}
\affiliation{School of Electrical Engineering and Telecommunications, University of New South Wales, Sydney, New South Wales 2052, Australia}
\author {Chris Adambukulam}
\affiliation{School of Electrical Engineering and Telecommunications, University of New South Wales, Sydney, New South Wales 2052, Australia}
\author {Arne Laucht}
\affiliation{School of Electrical Engineering and Telecommunications, University of New South Wales, Sydney, New South Wales 2052, Australia}
\author{Salvy P. Russo}
\affiliation{ARC Centre of Excellence in Exciton Science, School of Science, RMIT University, Melbourne, VIC 3001 Australia}
\affiliation{Chemical and Quantum Physics, School of Science, RMIT University, Melbourne VIC 3001, Australia}
\author{Alberto Peruzzo}
\email[Author to whom correspondence should be addressed: ]{alberto.peruzzo@rmit.edu.au}
\affiliation{Quantum Photonics Laboratory and Centre for Quantum Computation and Communication Technology, School of Engineering, RMIT University, Melbourne, Victoria 3000, Australia}

\begin{abstract}
Colour centres in diamond are attractive candidates for numerous quantum applications due to their good optical properties and long spin coherence times. They also provide access to the even longer coherence of hyperfine coupled nuclear spins in their environment. While the NV centre is well studied, both in experiment and theory, the hyperfine couplings in the more novel centres - SiV, GeV, and SnV - are still largely unknown.
Here we report on the first all-electron \textit{ab-initio} calculations of the hyperfine constants for SiV, GeV, and SnV defects in diamond, both for the respective defect atoms ($^{29}$Si, $^{73}$Ge, $^{117}$Sn, $^{119}$Sn), as well as for the surrounding $^{13}$C atoms. Furthermore, we calculate the nuclear quadrupole moments of the GeV defect.
We vary the Hartree-Fock mixing parameter for Perdew-Burke-Ernzerhof (PBE) exchange correlation functional and show that the hyperfine couplings of the defect atoms have a linear dependence on the mixing percentage. We calculate the inverse dielectric constant to predict an \textit{ab-initio} mixing percentage. The final hyperfine coupling predictions are close to the experimental values available in the literature. Our results will help to guide future novel experiments on these defects.
\end{abstract}

\maketitle

\section{\label{sec:introduction} Introduction}
Spins in the solid-state have long been proposed as a suitable platform for quantum information storage, processing, and networking~\cite{divincenzo2002spins,bennett2000quantum,divincenzo2000physical,DiVincenzoScience,Ruf2021}. 
Optically addressable spins in diamond show considerable promise as spin-photon interfaces for applications in quantum networks \cite{aharonovich2011diamond,simon2017towards}. This is a consequence of their ability to interface and entangle photons of optical wavelengths as flying qubits with spin qubits \cite{togan2010quantum, pfaff2014, bhaskar2020experimental}, as well as their milliseconds long coherence times \cite{sukachev2017silicon, herbschleb2019ultra} suitable for computational operations.

Whilst much of the past work on diamond was carried out with the nitrogen-vacancy centre, recently there has been growing interest in group-IV defects due to their improved photonic properties such as high brightness, indistinguishable emission of photons and spectral stability-- owing to the defects' D\textsubscript{3d} inversion symmetry that allows for sharp lifetime-broadened optical transitions and first order immunity to local electrical field fluctuations \cite{bradac2019quantum}. Furthermore, coherent control of group-IV defects \cite{pingault2017coherent,BeckerAllOptical,debroux2021quantum,siyushev2017optical,senkalla2023germanium,guo2023microwave}, has been demonstrated via magnetic and all-optical control techniques.
\begin{figure*}
    \centering    \includegraphics[width=0.8\textwidth]{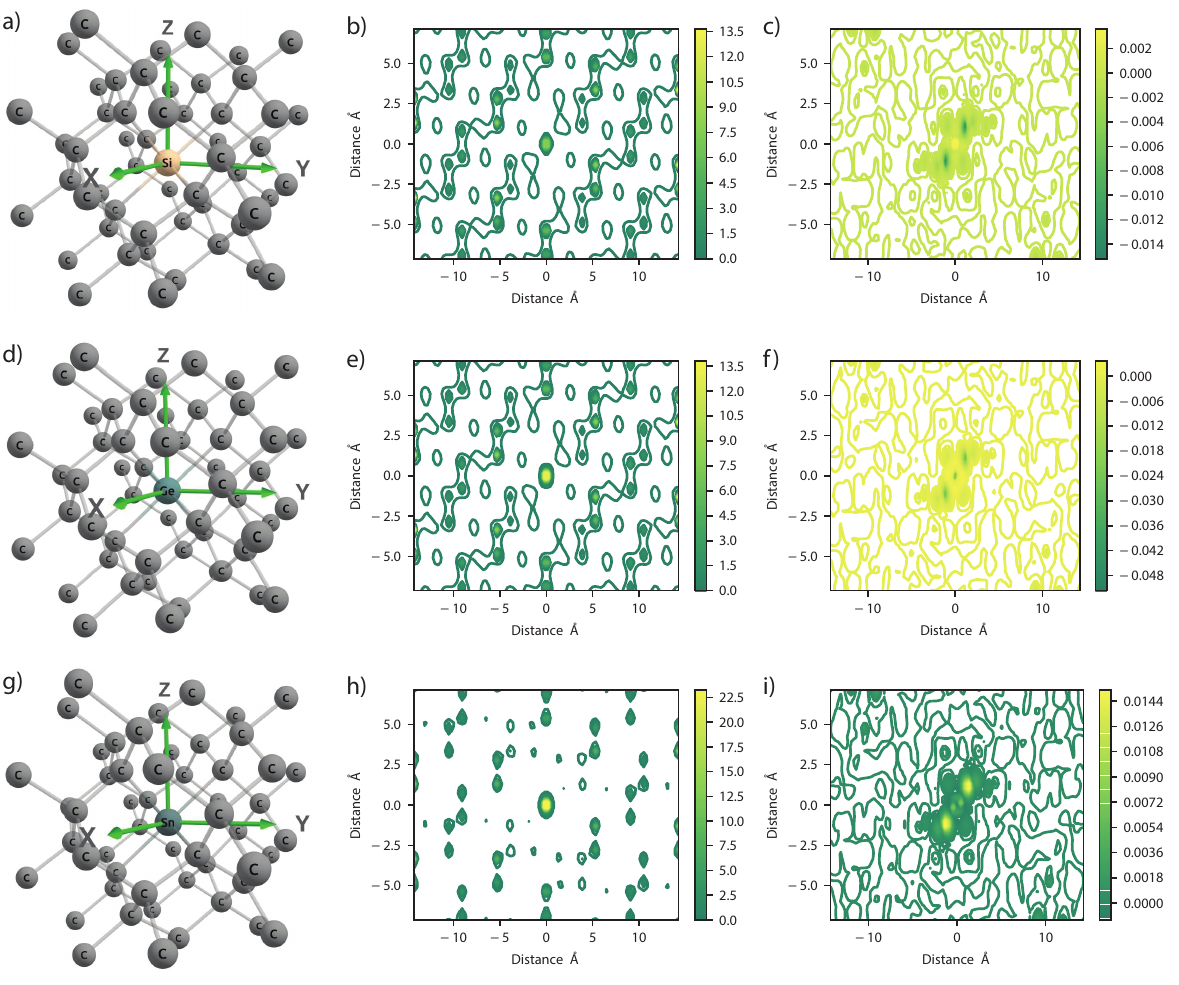}
    \caption{Optimised crystal geometry, electron density and spin density for (a-c) SiV$^-$, (d-f) GeV$^-$, and (g-i) SnV$^-$ defects in diamond. Electron and spin density are given in units of electrons per $\AA^3$ The plots of the charge and spin density are for the [111] crystallographic plane, which correspond to the plan parallel to Z and in between X and Y.}
    \label{fig:structures}
\end{figure*}

Nuclear spins within these defects may be even more suitable for quantum memory applications due to their robust decoupling from their environment \cite{fuchs2011quantum}, which naturally arises from their rather low gyromagnetic ratios (MHz/T). These nuclear spins are addressed via their hyperfine coupling to the defect electron spin. In essence, the electron spin acts as an ancilla with which nuclear spin initialization \cite{Chakraborty_2017}, readout \cite{Dutt2007}, control \cite{Zhang2019} and spin-photon entanglement~\cite{Tsurumoto2019} may be performed. These nuclear spins could either be randomly occurring \textsuperscript{13}C nuclei ($I=1/2$)   \cite{bradley,Zhang,Nguyen2019,Maity2022}, or the defect's intrinsic nuclear spin \cite{fuchs2011quantum,vallabhapurapu2022indirect,pingault2017coherent,parker2023diamond} which may be deterministically included by isotope selective implantation and provides a scalable alternative to the $^{13}$C spin. Moreover, there is potential for all-optical control of the intrinsic  nuclear spins of the group-IV defects\cite{BeckerAllOptical,vallabhapurapu2022indirect}. 
Such qualities make the intrinsic nuclear spins a valuable resource for applications in quantum networks and computation, especially when paired with the excellent optical qualities offered by the group-IV defects~\cite{bradac2019quantum}.

Remarkably, despite experimental progress and detailed theoretical investigations into the electronic structures of various group-IV defects, there is a lack of data and information on the nuclear spin isotopes for the group-IV defects in diamond. Particularly, current simulations to date have been able to predict the hyperfine coupling of the neighbouring atoms but not the intrinsic group-IV nuclear spin.
In this work, we address this by presenting all-electron \textit{ab-initio} simulation results that predict the hyperfine coupling strengths of the \siv, \sivminus, \gev, \gevminus, \snseven, \snsevenminus, \snnine and \snnineminus defects in diamond and the quadrupole moments of the $^{73}$GeV defects -- whose intrinsic nuclear spin is non-zero. We expect that our results will help to guide future novel experiments on these defects.

\section{Methodology and results}
The magnetic dipole of a nucleus and the spin of an electron interact via the hyperfine coupling. The strength of the interaction depends on direction, so a hyperfine tensor is defined that couples the spin operators of the nucleus and the electron. Explicitly, the hyperfine coupling Hamiltonian is,
\begin{equation}
H_{\rm hf} = \mathbf{S} A^N \mathbf{I},
\end{equation} where $\textbf{S} = (S_x, S_y, S_z)^T$ and $\textbf{I} = (I_x, I_y, I_z)^T$ are the vector electron and nuclear spin operators. As the magnetic dipoles of nuclei are known, only the spin density of the electron needs to be calculated. This can then be rewritten as two terms. The first is the interaction due to spin density at the nucleus site, which is isotropic and called the Fermi contact term.

\begin{figure*}
\centering
    \includegraphics[width=0.8\textwidth]{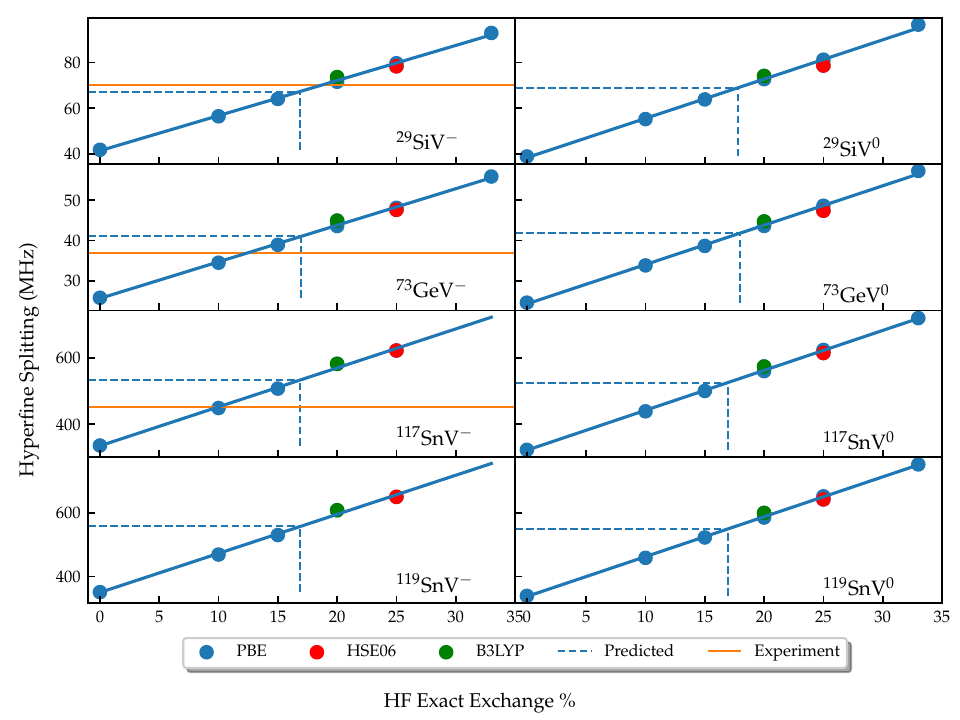}
\caption{Hyperfine coupling (total tensor in the z-direction) as a function of the Hartree-Fock exact exchange energy. }
\label{defect_hs}
\end{figure*}

The second is an anisotropic term due to dipolar interactions where the electron and nucleus are separated. In other words, for a nucleus $N$, we have:

\begin{equation}
    A^N_{i,j} = FC^N + T^N_{i,j},
\end{equation}

where $A^N_{i,j}$ is the full hyperfine tensor; $FC^N$ is the Fermi contact term; and $T^N_{i,j}$ is the dipolar term. Note that $i,j$ labels are the co-ordinates $x, y, z$.

Given $A^N$ is symmetric, we may also define a transformation matrix $R$ that transforms the coupling Hamiltonian to 
\begin{equation}
H'_{\rm hf} = A_x S_x I_x + A_y S_y I_y + A_z S_z I_z.
\end{equation} 
We provide $R$ in Tables S7-S12 in the Supplemental Material for all of the considered nuclear spins.

To correctly determine the electron spin density at the nucleus site, previous calculations based on pseudo potential methods \cite{Defo2021,harris2023}, required extra corrections. This corrective method has been also applied to calculate the hyperfine coupling to neighbouring \cthirteen. Furthermore, the effect of hybrid functionals was not explored~\cite{harris2023}.
Rather than using approximated pseudo-potential calculations, our method explicitly includes the all-electron basis set. 
In this work, we do not study the PbV defect center because it is challenging to find an appropriate all-electron basis set for this large system. Additionally, lead is a heavy element, which requires relativistic corrections to the core electrons. These corrections are computationally expensive and difficult to implement, so, in this study, we only consider the other three group-IV atoms (Si, Ge, Sn).

\begin{table*}[]
\begin{tabular}{|l|c|c|c|c|}
\hline
    \begin{tabular}[c]{@{}c@{}}  Color center \end{tabular} & 
    \begin{tabular}[c]{@{}c@{}}  Dielectric constant \end{tabular} &
    \begin{tabular}[c]{@{}c@{}}  HF mixing \% \end{tabular} &
        \begin{tabular}[c]{@{}c@{}} A$_{z}$ {[}MHz{]} \\(at mixing value) \end{tabular} &
    \begin{tabular}[c]{@{}c@{}}Experimental\\ A$_{z}$ {[}MHz{]} \end{tabular} \\ 
\hline
\siv     &5.61& 17.82  & 68.97 & --                                                                            \\
\sivminus     &5.94& 16.83  & 67.23 & $70\pm2$ \cite{Pingault2017}  \\
\gev     &5.56& 17.99  & 41.83 & --                                                                           \\
\gevminus     &5.90& 16.95 & 41.00 & $36.98\pm0.02$ \cite{adambukulam2023hyperfine}                                                                                 \\
\snseven &5.90& 16.94 & 525.43 & --                                                                            \\
\snsevenminus &5.92& 16.90 & 533.27 & $452\pm7$ \cite{parker2023diamond}                                                                           \\
\snnine   &5.90& 16.94 & 550.01 & --  \\
\snnineminus &5.92& 16.90 & 557.91 & --                                                                            \\
\hline
\end{tabular}
\caption{Predicted dielectric constants, HF mixing percentages and hyperfine coupling frequencies in the z direction.}
\label{table1}
\end{table*}

\begin{table*}[ht]
\begin{tabular}{l c c c c }
    \hline
    \hline
     Defect & $Q$~(kHz) & $\mathcal{V}_{xx}$~(nV/b) & $\mathcal{V}_{yy}$~(nV/b) & $\mathcal{V}_{yy}$~(nV/b)\\
     \hline\\
     %\gevminus & -0.1978 & -0.09584935 & -0.09584935 & 0.1916987\\
     %\gev & -0.2129 & -0.1030843 & -0.1030843 & 0.2061686\\
     \gevminus & $-183.9 \pm 0.9$ & -93.14027864 & -93.14027864 & 186.28055727 \\
     \gev      & $-198 \pm 1$ & -100.17074112 & -100.17074112 & 200.34148223 \\
    \hline
\end{tabular}
\caption{Quadrupole interaction strength and EFG for the GeV defect in the negative and neutral charge states.}
\label{table_quandruple_moment}
\end{table*}

The Fermi contact term is a magnetic interaction between an electron and an atomic nucleus. It is calculated by first evaluating the average spin density at the nucleus, $\langle\rho^{\rm spin}(r_{N})\rangle$, where $r_{N}$ is the position of the nucleus. The Fermi contact term in atomic units is then given by:

\begin{equation}
    \frac{1}{n_\alpha - n_\beta}  \frac{2}{3} \langle\rho^{\rm spin}(r_{N})\rangle,
    \label{eq-2}
\end{equation}

where $n_\alpha - n_\beta$ are the number of unpaired electrons.

The anisotropic term ($T^N_{i,j}$), in atomic units, is given by:

\begin{widetext}
\begin{equation}
     \frac{1}{n_\alpha - n_\beta} \sum_{\mu,\nu} \sum_g \rho^{\rm spin}_{\mu,\nu,g}
     \int \phi_\mu (r) \left( \frac{|r-r_N|^2\delta_{i,j} - 3(r-r_N)_i (r-r_N)_j}{{|r-r_N|^5}} \right)
     \phi^g_\nu(r) dr,
     \label{eq-3}
\end{equation}
\end{widetext}

where $(r - r_N)_i$ refers to the $i$th component of the vector. $\phi^g_{\nu}$ denotes the $\nu$th atomic basis function in cell $g$ in direct space and $\rho^{spin}_{\mu,\nu,g}$ is the element of the spin density matrix for the specified pair of orbitals. 

The spin density $\rho^{\rm spin}$ requires specifying the crystal structure and the exchange-correlation functionals.
We start by defining the crystal structure of a diamond cell containing the x-vacancy center, where x is Ge, Si, Sn. Using Crystal17 software \cite{crystal17} we specify the basis set as double zeta valence potential~\cite{Bredow2019,mike_towler_sn}, and exchange-correlation functional to be used in the calculation. 
Figure \ref{fig:structures} shows the optimised geometry of the crystal structure of the color centers under consideration. 

Here, PBE, HSE and B3LYP are used as exchange-correlation functionals \cite{rappoport2006approximate}. These are generalised gradient approximation (GGA) functionals mixed with a portion of the exact Hartree-Fock (HF) exchange energy. The hybrid functionals have been shown to provide accurate results for a wide range of molecular and solid-state systems, particularly for properties that are sensitive to electron correlation effects, such as reaction energies~\cite{Zhao2011}, redox potentials~\cite{Chevrier2010}, and ionization potentials and electron affinities~\cite{Miao2014}. However, the accuracy of the results obtained using them depends on the choice of the mixing parameter that determines the fraction of HF exchange energy included in the functional. This is well established for bandgap calculations \cite{Marques_2011}. Previously, this parameter has been found by fitting experimental data, e.g. in Ref.~\cite{Becke1993}.

For PBE we use several popular mixing values, 0\%, 10\% , 15\%, 25\% and 33\%. The results, reported in Fig. \ref{defect_hs}, show no dependence on the choice of GGA functional, but a strong linear dependence on the mixing parameter. Therefore, by performing a linear fit to the \sivminus defect data points, we are able to identify the mixing corresponding to the experimental hyperfine coupling A$_z$ of 70~MHz reported in Ref.~\cite{Pingault2017}, which results in 18.64\%. 
%table
Instead of relying on experimental data, it is preferable to predict the mixing percentage from \textit{ab-initio} methods. Previously, a procedure for inferring the Hartree-Fock mixing using the dielectric tensor has found success in predicting bandgaps~\cite{Koller_2013}. We now propose to use this method to determine the mixing required for hyperfine coupling. The dielectric tensor is computed using the coupled-perturbed HF calculation up to second order using Crystal17~\cite{crystal17,Ferrero2008-dc,Ferrero2008-ml,Ferrero2008-pw}, with PBE exchange correlation (no mixing).
The mixing parameter is given by $1/\epsilon^*_{\rm PBE}$, where $\epsilon^*_{\rm PBE}$ is the geometric mean of the diagonal elements of the dielectric tensor, reported in Table \ref{table1}. The mixing parameters are found in the range [16.83 - 17.99]. 
Next, we use these mixing values to calculate the A$_z$ component of the hyperfine tensor.
For \sivminus the mixing from the \textit{ab-initio} method is 16.83\%, which matches well the 18.64\% prediction from the fitting method.
Furthermore, recent experimental results for the \gevminus defect have reported a hyperfine value of $36.98 \pm 0.02$~MHz \cite{adambukulam2023hyperfine}. This compares favorably to our predicted value of $41~MHz$ using a calculated mixing parameter of $16.95\%$.
Furthermore, as expected, the calculated dielectric constants, also reported in Table \ref{table1}, are quite close to the value for bulk diamond of 5.7.

\begin{figure*}
\centering
\includegraphics[width=0.8\textwidth]{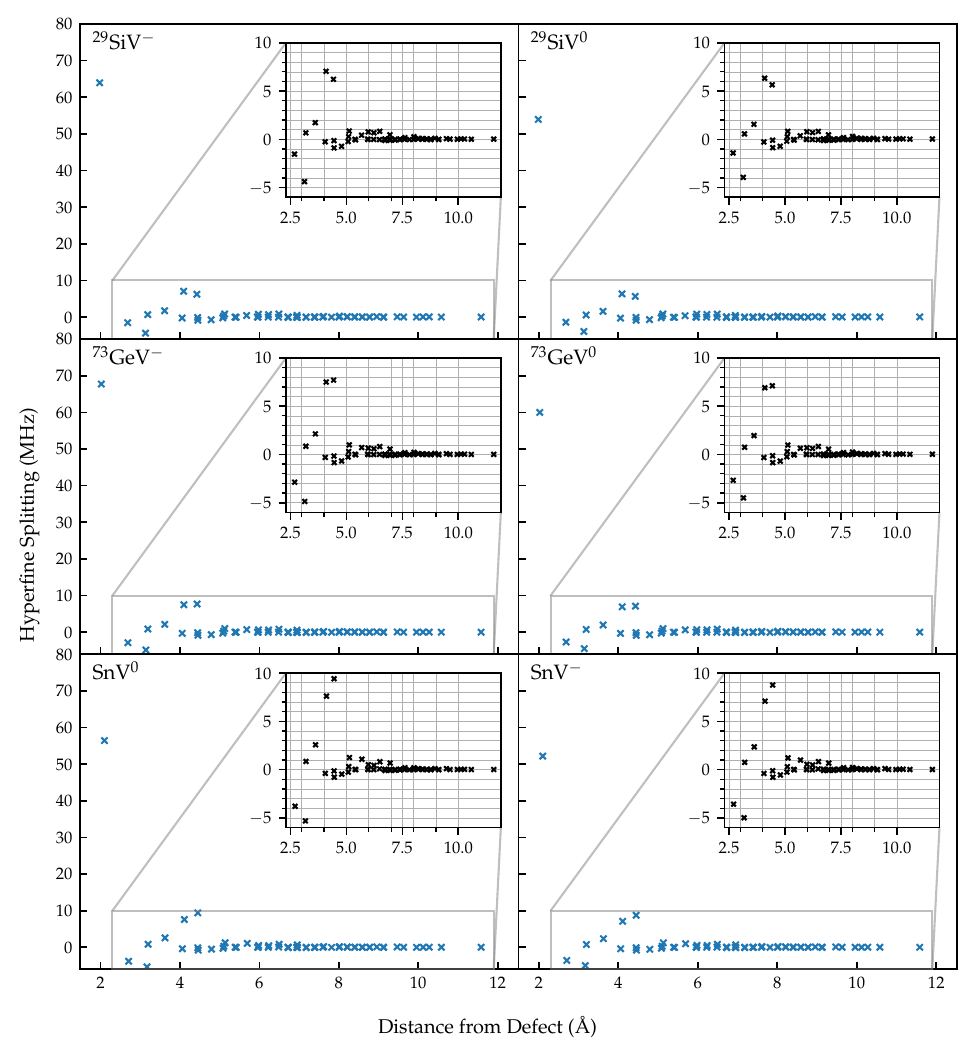}
\caption{Hyperfine coupling (Fermi contact) for the \textsuperscript{13}C atoms as a function of the distance from the defect.}
\label{carbons_hs}
\end{figure*}

We also investigate the effect of \cthirteen~ nuclei at different distances from the defect. This is an invaluable resource to help identify the source of hyperfine couplings in experimental investigations and to extend the use $^{13}$C nuclear spin registers ~\cite{bradley} from the NV center to the group-IV defects.
While the nearest neighbour carbon coupling has been reported in the literature \cite{Defo2021}, here we perform a calculation of up to 12 atomic units, corresponding to a 512 unit cell. 
For this we use the same procedure as above, but we set the value r$_N$ in equations \ref{eq-2} and \ref{eq-3} to be the site of the \cthirteen~ atom.
The results, reported in Figure \ref{carbons_hs} show that the \cthirteen~ as far as 4 \AA~can have non-negligible hyperfine coupling. Given in Tables S1-S6 in the Supplemental Material are the calculated values of the isotropic and anisotropic hyperfine constants for the $^{13}$C for each lattice site and for each of the color centres under study.

\subsection{Nuclear quadrupolar interaction of the Ge defect}

A nucleus with a spherically symmetric charge distribution can be described by the Coulomb potential of a point charge. However, if the charge distribution is not spherically symmetric, then additional terms are needed to describe the nucleus' electrical response to its environment. The lowest-order, non-zero correction term is the nuclear quadrupole interaction. This term captures deviations from spherical symmetry that are oblate (flattened) or prolate (elongated). It is a second-order moment, which means that it is sensitive to the second derivative of the charge distribution. More explicitly, the local electric field gradient (EFG) couples to the nuclear spin with the term, 
\begin{equation}
    H_{\rm{Q}} = \frac{1}{6} \frac{e q_n}{2I(2I-1)} \sum_{i,j} \mathcal{V}_{i,j} I_{i} I_{j},
\end{equation} where $e$ is the charge of an electron, $q_n$ is the quadrupole moment and $\mathcal{V}_{i, j}$ and $I_{i (j)}$ are EFG tensor components and nuclear spin operators over $i, j \in \{x, y, z\}$ \cite{asaad2020coherent}. As the EFG is a symmetric and trace-less tensor, there exists a set of principle axes $\{x', y', z'\}$ for which the tensor is diagonal. This reduces the coupling Hamiltonian to,
\begin{equation}
    H_{\rm{Q'}} = Q \bigg[I_{z'}^2 - \frac{I^2}{3} + \frac{\eta}{3} \big(I_{x'}^2 - I_{y'}^2 \big) \bigg].
\end{equation} where,
\begin{equation}
    Q = \frac{3 e q_n}{4I(2I-1)} \mathcal{V}_{z'z'} \,\, \rm{and} \,\, \eta = \frac{\mathcal{V}_{x'x'} - \mathcal{V}_{y'y'}}{\mathcal{V}_{z'z'}}.
\end{equation} We compute the EFG for the defect centers under study (see Table~\ref{table_quandruple_moment}), calculated for the PBE exchange correlation functional (no mixing). Due to symmetry considerations ($^{73}$Ge is the only isotope with a spin quantum number exceeding $1/2$), only \textsuperscript{73}Ge has a non-zero quadruple moment. Specifically, its quadrupole moment is $q_n = 0.196 \pm 0.001$~b~\cite{Kello1999}, where $1$~b $ = 10^{-28}$~m\textsuperscript{-2}. It turns out that for the GeV, $z'$ coincides with the defect high symmetry axis and the EFG tensor asymmetry is $\eta = 0$. Therefore, the quadrupolar interaction is entirely oriented along $[111]$ with interaction strengths of $Q = -183.9 \pm 0.9$~kHz and $Q = -198 \pm 1$~kHz for the \textsuperscript{73}GeV\textsuperscript{-} and \textsuperscript{73}GeV\textsuperscript{0}, respectively.

\section{Discussion and conclusion}

We have reported the first all-electron calculation of the hyperfine constants for \siv, \sivminus, \gev, \gevminus, \snseven, \snsevenminus, \snnine and \snnineminus defects in diamond. 

Using a fully \textit{ab-initio} approach, we have predicted the hyperfine couplings of the defect atom and showed that the hyperfine coupling has a linear dependence on the mixing percentage. This enables a linear fitting and extraction of mixing percentage from experimental data. It is notable that the HSE and B3LYP exchange-correlation functionals agree with PBE0 for their defined mixing values, 25\% and 20\% respectively. 
Furthermore we have shown that the actual mixing percentage can be found by an \textit{ab-initio} approach. Using this mixing percentage, the calculated hyperfine coupling is in good agreement with experimental values reported in the literature.

We have also performed hyperfine calculations for neighboring carbon nuclei beyond nearest neighbors.
Previous reports used pseudopotentials~\cite{Defo2021} and computed hyperfine coupling from neighboring carbons, concluding that the contribution of the exchange energy diminishes as the distance from the defect increases. We have considered a larger number of neighboring carbons and showed non-negligible contributions up to distance 4 \AA. 
Finally we have reported the nuclear quadrupole moments of the GeV defect.

These results can be useful to experimental investigations, providing a prediction of the hyperfine coupling and quadruple moment frequencies, which can guide the identification of the source of hyperfine coupling.

\section*{Acknowledgements}
A.P. acknowledges a RMIT University Vice-Chancellor’s Senior Research Fellowship, a ARC DECRA Fellowship (No. DE140101700), and a Google Faculty Research Award. This work was supported by the Australian Government through the Australian Research Council under the Centre of Excellence scheme (Nos. CE170100012 and CE170100026). It was also supported by computational resources provided by the Australian Government through the National Computational Infrastructure Facility.
H.H.V. acknowledges support from the Sydney Quantum Academy and the Australian Government Research Training Program Scholarship. C.A and A.L acknowledge support from the University of New South Wales Scientia program.

\bibliography{hyperfine_splitting_no_url}

\end{document}